\documentstyle[titlepage,12pt]{article}
\begin{document}
\baselineskip 4.5ex
\parindent 2em

\begin{titlepage}
\begin{center}
\vspace{12mm}
{\LARGE The high-field superconducting transition
induced by correlated disorder}
\vspace{25mm}

Zlatko Te\v sanovi\' c and Igor F. Herbut\\
{\sl Department of Physics and
Astronomy,  The Johns Hopkins University, Baltimore, MD 21218 }

\end{center}
\vspace{10mm}

\noindent
{\bf Abstract:}
The ``glassy'' superconducting transition at
high magnetic fields can
be induced by columnar disorder. A model is proposed in which
the thermodynamics of Bose condensation
of Cooper pairs into the lowest
Landau level eigenstate of the
random potential can be solved exactly. The
solution reflects a peculiar character of the high-field limit:
For example, the effective dimensionality of the transition
is shown to be a function of magnetic field.


\end{titlepage}

% pocetak
%\section{  }
The problem of superconducting transition
in the presence of strong
disorder is of both practical and
theoretical interest. Technologically,
the goal is to introduce defects into the sample in a way that
maximizes pinning of vortices and
increases critical currents \cite{1}.
The theoretical challenge is to
understand the mechanisms and the nature of
superconducting transition for various types of disorder. A
variety of novel phases have been
proposed, differing in the cases
of point like \cite{2,3} and
line-like disorder \cite{4,5}.
In this Letter we present a theory
of superconducting transition
at {\it high} magnetic fields ($> 1$ Tesla in HTS)
{\it induced} by the presence of columnar (line-like) defects.
In the absence of disorder, the high-field fluctuations
of the order parameter, $\psi (\vec r)$, are
strongly enhanced by formation
of Landau levels (LLs) for Cooper pairs
\cite{eilenberger}.
Such fluctuations lead to $D\to D-2$-dimensional reduction
in the pairing susceptibility, $\chi_{sc}(\vec r,\vec r')$,
and eliminate the superconducting
(Abrikosov) transition for $D=2,3$
\cite{dw}. The Abrikosov phase is then
replaced by a new fluctuation-induced state, the
density-wave of Cooper pairs (SCDW),
in which the thermal average $\langle
|\psi (\vec r)|^2\rangle$ has a weak modulation
accompanied by only a short range phase coherence \cite{dw},
\cite{footi}. In the presence of disorder
the LL degeneracy is lifted and
a possibility of superconducting
transition is restored. $\chi_{sc}$
can now diverge at some finite
temperature, $T_{sc}(H)$, determined
by the strength of disorder \cite{dw}. At $T_{sc}(H)$
the normal state is unstable to
Bose condensation of Cooper pairs
into the lowest energy eigenstate
of the random potential, which we
argue extends over the whole
sample in situations of experimental interest.
Furthermore, for experimentally relevant parameters,
$T_{sc}(H)$ can be far above the SCDW
transition line over much of
the $H-T$ phase diagram, allowing us
to treat the correlations
that produce SCDW in an approximate way.

We consider a realistic
model for a superconductor in a magnetic field parallel
to the columns which is exactly solvable.
The model exhibits ``dimensional transmutation",
{\it i.e.} the effective dimensionality
of the transition changes
continuously as a function of magnetic field. This effect
is a direct consequence of analytic properties of
LL wavefunctions and is a signature of the high-field limit.
We determine the transition line in $H-T$ phase
diagram, the Edwards-Anderson order
parameter, and the behavior of correlation length,
specific heat and magnetic susceptibility in
the vicinity of the transition.
There are similarities between the transition
considered here and the one in
the spherical model for spin-glasses \cite{6}.

We are interested in strongly
anisotropic layered superconductors
described by the Ginzburg-Landau (GL) Lawrence-Doniach
model, with magnetic field perpendicular to the layers.
Fluctuations of the magnetic field
are neglected ($\kappa \gg 1$). We
focus on the {\it high-field limit},
where the LL structure of Cooper pairs
dominates the fluctuation spectrum:
This is the case for fields above
$H_{b} \approx (\theta/16)H_{c2}(0)(T/T_{c}(0))$,
where $\theta$ is the
Ginzburg fluctuation parameter \cite{7}.
(For instance, in BSCCO 2:2:1:2,
$\theta\approx 0.045$ and $H_{b}\approx 1$ Tesla.)
In this regime,
the essential features of the
physics are captured by retaining only
the lowest Landau level (LLL) modes. This
is the renormalized GL-LLL theory \cite{dw,7}.
The partition function
is $Z=\int D[\psi^{*} \psi] \exp(-S)$, and
\begin{equation}
S=\frac{d}{T} \sum_{n=1} ^{N_{L}}\int d^{2}\vec{r}
\left\{ \eta |\psi_{n}(\vec{r})-
\psi_{n+1}
(\vec{r})|^{2} + \left[\alpha '(T,H)+
\lambda \sum_{i} \delta (\vec{r}-\vec{r_{i}})\right]
|\psi_{n}(\vec{r})|^{2} + \frac{\beta}{2}
|\psi_{n}(\vec{r})|^{4}\right\},
\end{equation}
 where $\alpha '(T,H) =a(T-T_{c2}(H))$, $d$
is the effective layer separation,
$n$ is the layer index,
$\lambda>0$ is the effective strength
of the defects and $a$, $\beta$ and
$\eta$ are phenomenological parameters.
The magnetic field is
assumed to be parallel to columnar defects,
the effective potentials of all
defects the same and well-represented by delta-functions.
Random variables in the problem are 2D coordinates of
defects, $\{\vec{r_{i}}\}$. We assume that columns of
damaged superconductor are distributed according to
the Poisson distribution
$P_{N}(\vec{r_{1}},...\vec{r_{N}})=
(e^{-\rho A} \rho^{N})/N! $
where $P_{N}$ is the probability
for finding $N$ impurities at the
positions $\vec{r_{1}},...\vec{r_{N}}$,
$A$ is the area of the system
and $\rho$ is the concentration of impurities \cite{8}.
After rescaling the fields
and the lengths as $(2d\beta 2\pi l^{2}/T)^{1/4}
\psi \rightarrow \psi$,
$r/(l\sqrt{2 \pi})\rightarrow r$,
where $l$ is the magnetic length
for charge $2e$, the quartic term can be rewritten
\begin {equation}
\frac{1}{4}\sum_{n}\int d^{2}\vec{r} |\psi_{n}|^{4} =
\frac{1}{4N}\sum_{n} \beta_{A}(n)
(\int d^{2}\vec{r} |\psi_{n}|^{2})^{2}
\end{equation}
where $N=A/2\pi l^{2}$ is the degeneracy of the LLL
and $\beta_{A}(n)=(N \int |\psi_{n}|
^4)/(\int |\psi_{n}|^{2})^{2}$
is the generalized Abrikosov ratio
corresponding to configuration
$\psi_{n}(\vec r)$. We now observe that $\beta_A (n)$
is only weakly dependent on
the actual configuration, the well known
example being the small difference in $\beta_{A}$ between
triangular and square lattice of zeroes \cite{7}.
Thus, we may substitute
$\beta_{A}(n)$ in the quartic term by
its thermal average, $\langle\beta_{A}\rangle$,
and treat this as an input to
the theory. This approximation neglects
the non-perturbative lateral
correlations that produce the SCDW
transition \cite{dw}. It is justified
if $T_{sc}(H)$ is far above the
SCDW transition line. In that case
the SCDW correlations enter
only very close to the transition
and can be ignored in most
realistic situations. Since
superconducting and SCDW transitions
arise from two distinct mechanisms,
the respective transition lines
scale differently in the
$H-T$ phase diagram and, for moderate
disorder, we are assured of a
wide region near $H_{c2}(T)$
where the neglect of SCDW correlations should be
justified (see Fig. 1) \cite{footii}.

After the $\beta_A (n)\to \langle\beta_A\rangle$
substitution the
thermodynamics of the model becomes exactly
solvable. We first introduce
variables $\{x_{n}\}$ to decouple the quartic term
and integrate over the fields $(\psi^{*},\psi)$. This leads to
$Z=\int \prod_{n} dx_{n}\exp(-NS')$, where
\begin{equation}
S'= -\sum_{n} \frac{x_{n}^{2}}{\langle \beta_{A}\rangle} +
 \int_{0}^{\infty} dV \rho_{f}(V)
Tr_{(n,m)} \ln {\left[g_{\eta}(2\delta_{n,m}-
\delta_{n,m-1}-\delta_{n,m+1})+
(g_{\alpha}+x_{n}+g_{\lambda}V)\delta_{n,n}\right]}.
\end{equation}
We drop terms coming from the
rescaling of $\psi_n (\vec r)$ and
introduce dimensionless combinations of
GL parameters $g_{\eta,\alpha,\lambda}=
\{\eta,\alpha ',\lambda/2\pi l^{2} \}\times\sqrt{
(d\pi l^{2})/(T\beta)}$.  The density of states
for a disordered system in the LLL can be found
exactly by using the supersymmetric
formalism \cite{9}. For
the Poisson short-range
scatterers it is given by
\begin{equation}
\rho_{f}(V)=\frac{1}{\pi}Im\frac{d}{dV}ln
\int_{0}^{\infty} dt \exp(iVt-
f\int_{0}^{t}\frac{dy}{y}(1-e^{-iy}))
\label{4}
\end{equation}
where $f=\rho 2\pi l^{2}$.
 In the thermodynamic limit $N\rightarrow\infty$,
 the partition function
is completely determined by the saddle-point of $S'$.
Assuming that the saddle
point is at $x$ independent of the
layer index, we finally write
the free energy above the critical
temperature
\begin{equation}
\frac{F}{NN_{L}T}=-\frac{x^{2}}{\langle \beta_{A}\rangle}
+\frac{1}{2} \int_{-1}^{1} dk
\int_{0}^{\infty}dV\rho_{f}(V)\ln \left[g_{\eta}e(k)+
g_{\alpha}+x+g_{\lambda}V\right]~~~,
\end{equation}
where $e(k)=1-\cos (k)$ and $x$
is determined by the solution of
\begin{equation}
x=\frac{\langle\beta_{A}\rangle}{4}\int_{-1}^{1}
dk \int_{0}^{\infty} \frac{\rho_{f}(V) dV}
{g_{\eta}e(k)+g_{\alpha}+x+g_{\lambda}V}~~~.
\label{6}
\end{equation}

In Eq. 6 it is important to know
the behavior of density of
states at low energies. For $f<1$,
density of states has a delta-function
singularity at $V=0$, whilst for $f>1$,
$\rho_{f}(V)\sim V^{f-2}$
when $V\rightarrow 0$ \cite{9}. When $V<0$, $\rho(V)\equiv 0$,
as also can be inferred from Eq. 4 \cite{9}.
The transition line, $T_{sc}(H)$, in the
$H-T$ diagram is determined by  Eq. 6 and
$x+g_{\alpha}=0$, which corresponds
to condensation of Cooper pairs
into $k=0$ and $V=0$ eigenstate of the random potential.
It is easily seen that
there will be a non-zero transition
temperature only if concentration
of impurities and magnetic field are
such that $f>3/2$. Below this value of $f$
LLL degeneracy  is not sufficiently lifted by the random
potential and thermal fluctuations
prevent a finite temperature phase transition
in our model \cite{footiii}. Experimentally,
this should manifest itself as
a drop in transition temperature
when the field exceeds a certain value.  $f=3/2$
determines the effective
lower critical dimension for our model.
After introducing dimensionless quantities $t=T/T_{c}(0)$,
$h=H/H_{c2}(0)$ and
$\lambda '=\lambda H_{c2}(0)/\phi_{0}aT_{0}$,
where $\phi_{0}$ is the flux quantum,
we perform the integration over wave-vector
$k$ in Eq. 6 to obtain the expression
for transition temperature
\begin{equation}
t_{sc}(h)=(1-h) \left[1+\frac{\theta\langle
\beta_{A}\rangle}{2\lambda '}
\int_{0}^{\infty}\frac{\rho_{f}(V) dV}
{\sqrt{V^{2}+(2\eta V)/(h\lambda '
aT_{0})}}\right]^{-1}.
\end{equation}

Notice that when $\lambda '\rightarrow 0$
we have $t_{sc}(h)\rightarrow 0$,
while for $\lambda '\rightarrow \infty$,
$t_{sc}(h)\rightarrow 1-h$
\cite{footiv}.
Also, with increasing parameter $f$,
$t_{sc}(h)$ increases. This is
related to the observation in Ref. 1
that the irreversibility line shifts
to higher temperatures with increasing
doses of irradiation with heavy ions.
Numerical solution for $t_{sc}(h)$ is
displayed in Fig. 1 for $\lambda '=1$,
$\eta/aT_{0}=.01$, $\theta=0.03$,
$\langle \beta_{A}\rangle=1.3$
and $f=0.04/h$.
We have set $\langle \beta_{A}\rangle$
to a constant for simplicity. If
$H_{c2}(0)\approx 100$ Tesla, $f=0.04/h$
corresponds to average distance between
defects of $225$\AA~(at 1 Tesla, $l\cong 180$\AA).
The diameter of the columns
depends on the size and energy
of particles used for irradiation but it is about
$50$\AA~and hence much smaller than
the magnetic length for the fields of
interest.  Thus, representing defects
by delta-functions is appropriate.

As temperature drops below $t_{sc}(h)$, $x$ remains
at the value it had at the
transition. There is now macroscopic occupancy
of the lowest energy state at $V=0$ and $k=0$.
As is well known, condensation
into this state is possible only if
the state is {\it extended}. It is a
special feature of this problem that this indeed
is the case for certain range
of impurity concentration. The density of
states, Eq. 4, changes at $V=0$ from being
infinite when $f<2$, to being
zero when $f>2$. Thus, for fields and
impurity concentrations such
that parameter $f<2$, true extended states
(which always exist in the LLL \cite{10}) must
lie at the bottom of the impurity band,
since the number of states there
diverges. The change of behavior in the density of
states at $f=2$ could be caused by the fact that
the mobility edge shifts to
positive energies at some $f_0>2$,
leaving spread-out but localized states
at $V=0$, which now becomes the tail of the distribution.
Numerical diagonalization studies indicate
that mobility edge is indeed located
near the band center for $f>4$ \cite{11}.
Thus, strictly speaking, our model is appropriate for
$f<f_0$. However, even for $f$ above but close to $f_0$,
which is often the case for fields and concentrations
of experimental interest, the
states at $V=0$ are still
near mobility edge and will appear
extended in a finite size sample.
On this basis, we expect that
useful information about the
transition can still be obtained
within our model.

With these cautionary remarks in mind,
the natural order parameter is the thermal average
of the component of $\psi_n(\vec r)$
corresponding to the eigenvalue with
$V=0$ and $k=0$. This is
$\langle \psi_{0,0} \rangle =
(NN_{L}(g_{\alpha}|_{t=t_{sc}(h)}-g_{\alpha})/
\langle\beta_{A}\rangle)^{1/2}$.
The disorder average value of the field is
$\overline{\langle\psi_{n}(\vec{r})\rangle}
=\sum_{V,k}
\overline{ \phi_{V}(\vec{r})
}\exp{(ikn)}\langle \psi_{V,k}\rangle=0$,
due to random phases of the
state $\phi_{V=0}(\vec{r})$.
Under the assumption that the lowest
state is extended through the sample,
$\overline{|\phi_{V=0}(\vec{r})|^{2}}\approx 1/N$;
the Edwards-Anderson order parameter \cite{12}
$q_{EA}=\overline{|\langle\psi_{n}(\vec{r})
\rangle|^{2}}$ then equals
\begin{equation}
q_{EA}=\frac{2}{\langle\beta_{A}\rangle}
(g_{\alpha}|_{t=t_{sc}(h)}-g_{\alpha})
\end{equation}
below $t_{sc}(H)$, and is zero above. Thus,
$q_{EA}\propto (t_{sc}(h)-t)^{2\beta}$,
with the exponent $\beta=1/2$.
The free energy below  $t_{sc}(h)$ is
\begin{equation}
\frac{F}{NN_{L}T}=-\frac{g_{\alpha}^{2}}
{\langle\beta_{A}\rangle} + \frac{1}{2}
\int_{-1}^{1}dk\int_{0}^{\infty}\rho_{f}(V)
dV \ln {(g_{\eta}e(k)
+g_{\lambda} V)}.
\end{equation}

To calculate the exponents that determine the
divergence of correlation
lengths parallel and perpendicular
to the field  we first note that
from Eq. 6 and the definition
of the critical line it follows
\begin{equation}
(g_{\alpha}+x)\left[1+\frac{\langle\beta_{A}\rangle}{4}
\int_{-1}^{1}dk\int_{0}^{\infty}
\frac{\rho_{f}(V)dV}{(g_{\eta}e(k)+g_{\lambda}V)(g_{\eta}
e(k)+g_{\lambda}V+g_{\alpha}+x)}\right]
=g_{\alpha}-g_{\alpha}|_{t=t_{sc}(h)}.
\end{equation}
The integral in the last equation diverges for $f<5/2$
as $(g_{\alpha}+x)^{f-5/2}$
when the transition line is approached from above,
and it is finite for $f>5/2$. Thus, we obtain
$(g_{\alpha}+x)\propto (t-t_{sc})^{1/(f-3/2)}$
for $f<5/2$ and
$(g_{\alpha}+x)\propto (t-t_{sc})$
for $f \ge 5/2$. The same
behavior follows if the transition
line is approached along the line of
constant temperature. This determines the
value of the exponent
$\nu_{\|}=1/(2f-3)$ for $f<5/2$ and the
classical value $\nu_{\|}=1/2$ for $f> 5/2$,
where the correlation
length parallel to the field is $\xi_{\|}
\propto [t-t_{sc}(h)]^{-\nu_{\|}}$.
The concentration corresponding to
$f=5/2$ determines the effective
upper critical dimension in the problem.
We now turn to the correlation
length perpendicular to the field,
$\xi_{\bot}\propto[t-t_{sc}(h)]^{-\nu_{\bot}}$
and study the susceptibility
associated with Edwards-Anderson order parameter,
\begin{equation}
\chi_{EA}(\vec{r}-\vec{r'})\equiv
\overline{\langle\psi_{n}^{*}(\vec{r})\psi_{n}
(\vec{r'})\rangle\langle\psi_{n}(\vec{r})
\psi_{n}^{*}(\vec{r'})\rangle}~~~.
\end{equation}
After expanding the field operators in the eigenbasis of
random potential we obtain
\begin{equation}
\chi_{EA}(\vec{r}-\vec{r'})=
\int \frac{dV_{1}dV_{2}dk_{1}dk_{2}
F(\vec{r}-\vec{r'},V_{1},V_{2})}
{(g_{\eta}e(k_{1})+g_{\lambda}V_{1}+
g_{\alpha}+x)(g_{\eta}e(k_{2})+
g_{\lambda}V_{2}+g_{\alpha}+x)}
\label{12}
\end{equation}
where the function $F$ is the
two-particle spectral density \cite{13}
\begin{equation}
F(\vec{r}-\vec{r'},V_{1},V_{2})=
\overline{\sum_{i,j} \delta(V_{1}-V_{i})
\delta(V_{2}-V_{j}) \phi_{i}^{*}(\vec{r})
\phi_{i}(\vec{r'})
\phi_{j}(\vec{r})\phi_{j}^{*}(\vec{r'})}
\end{equation}
and $\phi_{i}(\vec{r})$ are the
eigenstates of the random potential.
If we now introduce $V=(V_{1}+V_{2})/2$
and $\omega=(V_{1}-V_{2})/2$, for
$V$ close to the mobility edge and small $(q,\omega)$, the
Fourier transform of $F$
has a diffusive form \cite{13,14}
\begin{equation}
F(\vec{q},V,\omega)=\frac{\rho_{f}(V) q^{2} D(q^2/\omega)}
{\pi(\omega^{2}+q^{4}D^{2}(q^{2}/\omega))}~~~,
\end{equation}
where $D(q^{2}/\omega)$ is the  generalized ``diffusion constant".
Assuming this form for $F(\vec{q},V,\omega)$ and rescaling
everything by the appropriate power of temperature in
Eq. 12, we obtain $\nu_{\bot}=\nu_{\|}$. Note that
the density of states $\rho_{f}(V)$
is roughly constant except in a narrow
region, typically 1\% of total bandwidth,
around $V=0$, where it either diverges or vanishes.
Hence, unless one
experimentally probes the system very close to the transition,
the observed correlation length exponent would be
the one corresponding to $f=2$,
{\it  i.e.} $\nu_{\bot}=\nu_{||}=1$. This agrees well with
experimental results of Ref. 5.

Magnetization per unit volume equals
\begin{equation}
\frac{M}{AN_{L}d}=-\frac{2T_{0}\sqrt{th}}
{d\phi_{0}\sqrt{\theta}}\left(q_{EA}+
\frac{2x}{\langle\beta_{A}\rangle}\right)~~~.
\end{equation}
Below the transition line this
coincides with the usual mean-field result.
Above the transition line $q_{EA}=0$ and from Eq. 10
it follows that at constant temperature close to the transition
$(g_{\alpha}+x) \propto[h-h_{sc}(t)]^{1/(f-3/2)}$ when
$f<5/2$ and $(g_{\alpha}+x) \propto
[h-h_{sc}(t)]$ otherwise. Thus the magnetic
susceptibility is a smooth function
of the field at the transition
for $f<2$, but has an upward  cusp
for $2<f<5/2$ and a
discontinuity for $f>5/2$. The size of this discontinuity
depends on the location
of the transition in the $H-T$ diagram.
Differentiating
the free energy twice with respect to temperature
one obtains the
specific heat. It is straightforward to show that
at the transition it behaves the
same way as susceptibility; smooth for
$f<2$, has a cusp
for $2<f<5/2$ and has the usual discontinuity for
$f>5/2$.
More precisely, both magnetic susceptibility and specific
heat behave as $[t-t_{sc}(h)]^{-\alpha}$
for $3/2<f<5/2$, where $\alpha=(f-5/2)/(f-3/2)$.
The behavior of the specific heat, order parameter and
correlation length in our model is analogous to the
one obtained from O(2$N$) vector model in the limit
$N\rightarrow \infty$
and in the {\it effective} dimension $D_{eff}=2f-1$
\cite{15}. The magnetic susceptibility however,
behaves differently at the transition;
whilst it diverges in the O(2$N$) vector model with the
exponent $\gamma=[(D_{eff}/2)-1]^{-1}$, it is finite in our case
even below the effective
upper critical dimension. This is a consequence
of a diamagnetic nature of magnetization in our problem.

We should stress again that the critical behavior of our model
does not describe ``true" critical properties of the
GL-LLL theory with disorder, since we have
ignored lateral SCDW
correlations. Such correlations will always become
important sufficiently close to the transition. However, as it
is clear from Fig. 1, there is a wide region in the $H-T$
phase diagram where the superconducting transition lies
far above the SCDW transition line for clean systems. In this
region, the ``true" critical behavior will set in only
very near the $T_{sc}(H)$ line and our model should be
appropriate in most experimental situations.

In summary, we have studied the high-field
superconducting glassy transition induced by
columnar disorder. Using the exact density
of states for random array of short-range scatterers in the
LLL level and the assumption that the lowest eigenstate
of such a potential is extended over a finite size sample
under certain conditions, we have
obtained the Edwards-Anderson
order parameter, correlation length,
magnetization and the specific heat close to the transition.
The transition line in $H-T$ phase
diagram has also been calculated.
The critical exponents are found to depend on magnetic field.

This work
has been supported in part by the David and Lucile Packard
Foundation.

\pagebreak

Caption:

Figure 1: The $H-T$ phase
diagram for strongly type-II superconductor
with columnar disorder ($h\equiv H/H_{c2}(0)$,
$t\equiv T/T_{c0}$).
The full line represents the second-order phase
transition boundary
between normal and ``glassy" superconducting state
for the set of parameters given in the text. Dashed-dotted
line is the SCDW transition in clean system.
Dashed line is the mean-field $h_{c2}(t)$.  The LL approximation
breaks down in the shaded region at the bottom.

\pagebreak


\begin{thebibliography}{99}
\bibitem{footi}
This is the scenario expected in strongly fluctuating
layered systems, like
high-temperature superconductors (HTS), at high fields. In more
conventional isotropic systems the SCDW transition
could be sufficiently
strongly first order to produce a rather long range phase
coherence, extending over an entire sample \cite{dw}.
\bibitem{footii}
The SCDW transition is driven by the quartic term in the GL-LLL
functional while the superconducting transition arises chiefly
from the quadratic term. Consequently, the scaling of transition
temperature involves $\alpha '^2/\beta$ for the former
and $\alpha 'V/\beta$ for the latter, where $V$ characterizes the
strength of disorder.
\bibitem{footiv}
Of course, in the limit of very strong disorder the whole LL
structure may eventually
collapse. Thus, while we obtain sensible
results in the limit of very strong disorder, our model is best
suited for weak to moderate
disorder: Just when the disorder becomes
too strong is the question which should be settled empirically,
by comparing our results to experiments and other theoretical
models.
\end{thebibliography}
\end{document}